\documentclass[prl,amsmath,amssymb,twocolumn,superscriptaddress]{revtex4}
\usepackage{amsmath,amssymb}
\usepackage[usenames]{color}
\usepackage{amssymb}
\usepackage{grffile}
\usepackage[pdftex]{graphicx}
\usepackage{amsmath, amstext, amssymb, amsfonts, amsxtra}
\usepackage{textcomp}
\usepackage{xspace}
\usepackage{color}

\begin{document}
\title{One-dimensional quasicrystals with power-law hopping}
\author{X. Deng}
\affiliation{Institut f\"ur Theoretische Physik, Leibniz Universit\"at Hannover, Appelstr. 2, 30167 Hannover, Germany}
\author{S. Ray}
\affiliation{Indian Institute of Science Education and Research, Kolkata, Mohanpur, Nadia 741246, India}
\author{S. Sinha}
\affiliation{Indian Institute of Science Education and Research, Kolkata, Mohanpur, Nadia 741246, India}
\author{G. V. Shlyapnikov}
\affiliation{LPTMS, CNRS, Universite Paris Sud, Universite Paris-Saclay, Orsay 91405, France}
\affiliation{SPEC, CEA, CNRS, Universite Paris-Saclay, CEA Saclay, Gif sur Yvette 91191, France}
\affiliation{Russian Quantum Center, Skolkovo, Moscow 143025, Russia}
\affiliation{\mbox{Van der Waals-Zeeman Institute, University of Amsterdam, Science Park 904, 1098 XH Amsterdam, The Netherlands}}
\affiliation{Wuhan Institute of Physics and Mathematics, Chinese Academy of Sciences, 430071 Wuhan, China}
\author{L. Santos}
\affiliation{Institut f\"ur Theoretische Physik, Leibniz Universit\"at Hannover, Appelstr. 2, 30167 Hannover, Germany}

\date{\today}

\begin{abstract}
One-dimensional quasi-periodic systems with power-law hopping, $1/r^a$, differ  
from both the standard Aubry-Azbel-Harper (AAH) model and from power-law systems with uncorrelated disorder. 
Whereas in the AAH model all single-particle states undergo a transition from ergodic to localized at a critical quasi-disorder strength, short-range power-law hops with $a>1$ can result in mobility edges. We find that there is no localization for long-range hops with $a\leq 1$, in contrast to the case of uncorrelated disorder. Systems with long-range hops rather present ergodic-to-multifractal edges and a phase transition from ergodic to multifractal~(extended but non-ergodic) states. 
Both mobility and ergodic-to-multifractal edges may be clearly revealed in experiments on expansion dynamics.
\end{abstract}

\maketitle



\begin{figure}[t]
\begin{center}
\includegraphics[clip=true,width =1\columnwidth]{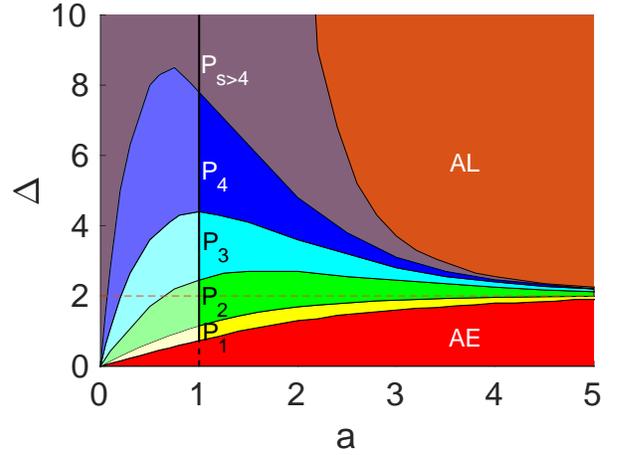}
\vspace*{-0.5cm}
\caption{(Color online) Regimes of 1D quasicrystals with power-law hopping, for $\beta=(\sqrt{5}-1)/2$. For small quasi-disorder strength $\Delta$ all SPS are ergodic~(AE) and for large 
$\Delta$ (for hopping power $a>1$) all are localized~(AL). The $P_s$ regimes are characterized by a fraction $\beta^s$ of ergodic SPS, 
whereas the rest are localized~($a>1$) or multifractal~($a\leq 1$).  The different behavior for $a>1$ and $a<1$ is indicated in the figure with a slightly different color. 
The results were obtained for $987$ sites, with periodic boundary conditions. Calculations for larger systems do not modify the results~\cite{footnote-SM}.
}
\vspace{-0.8cm}
\label{fig:1}
\end{center}
\end{figure}




\begin{figure*}[t]
\begin{center}
\includegraphics[width =2\columnwidth]{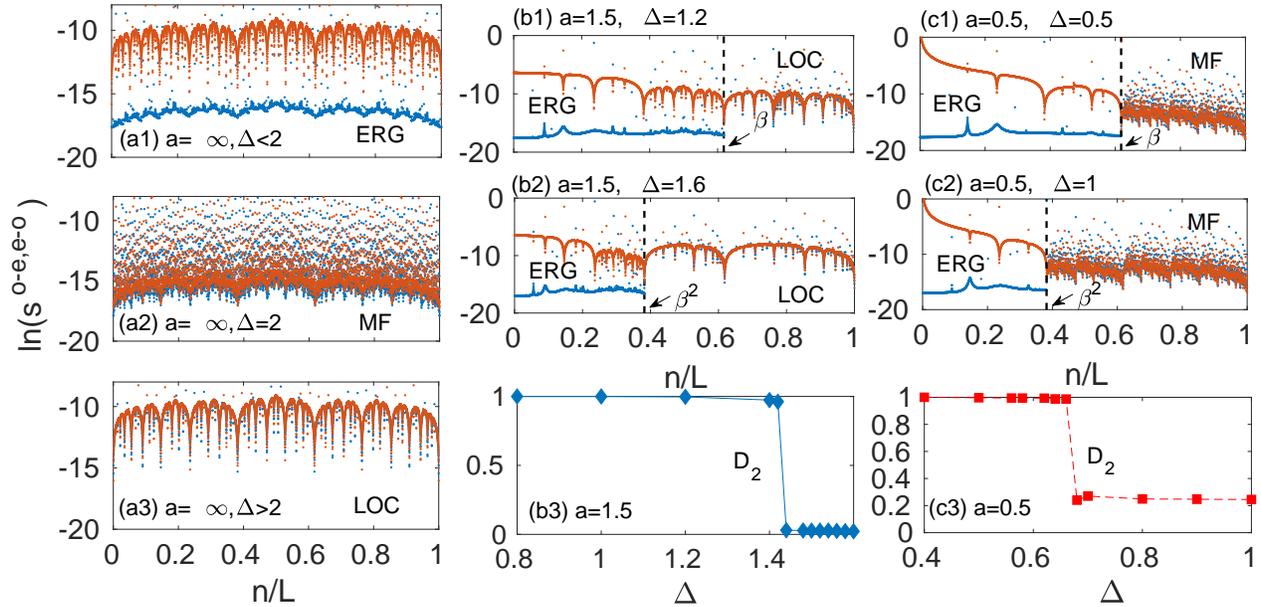}
\caption{(Color online) Results for $\beta=(\sqrt{5}-1)/2$. Level spacing $s^{e-o}$~(red) and $s^{o-e}~$(blue) for the AAH model~(a1-a3), $a=1.5$~(b1,b2) and $a=0.5$~(c1,c2) for 
different $\Delta$. In the AAH model all SPS are either localized~(LOC), multifractal~(MF) or ergodic~(ERG). In the GAAH model, $P_s$ regimes appear, in which the lowest $\beta^s$ fraction of SPS is ergodic, whereas the rest is localized ($a>1$) or multifractal ($a<1$).  These graphs were obtained from calculations for $L=28657$ sites with periodic boundary conditions. 
(b3) and (c3) show $D_2$ for the SPS between $\beta^2 L$ and $\beta L$ at $a=1.5$ and $0.5$, respectively. For $a=1.5$~($0.5$) a block-like localization~(ergodic-to-multifractal) transition occurs when crossing from $P_1$ to $P_2$. The results of panels (b3) and (c3) were obtained from calculations with up to $L=75025$ sites with periodic boundary conditions and then extrapolated to infinite systems. See Ref.~\cite{footnote-SM} for more details about the energy and $\Delta$ dependence of $D_2$.}
\vspace{-0.5cm}
\label{fig:2}
\end{center}
\end{figure*}

 
 
Quasicrystals constitute an intriguing intermediate case between disordered and periodic systems. In the former case arbitrarily small disorder 
results in localization for all single-particle states (SPS) in both one- and two-dimensional (1D and 2D) systems, whereas in three dimensions a mobility edge separates extended and localized SPSs~\cite{Anderson1958,Abrahams1979}. The situation is very different in quasi-periodic systems 
formed by two incommensurate lattices, which for 1D in the tight-binding regime~(with nearest-neighbor hopping) are well described by the  Aubry-Azbel-Harper~(AAH) model~\cite{Harper1955, Azbel1964, Aubry1980}. This model has been realized in experiments with ultracold atoms in bichromatic optical lattices, in which single-particle localization, Bose glasses, and many-body localization have been observed~\cite{Roati2008,Deissler2010, Schreiber2015,Lueschen2017}. Due to the self-duality of the AAH model~\cite{Aubry1980, Suslov1982, Wilkinson1984}, above a critical quasi-disorder strength all SPS change from ergodic to localized.

In disordered systems extended states were commonly believed to be ergodic, except at the mobility edge, where the states are 
multifractal, i.e. neither localized nor ergodic~\cite{Wegner1980,Aoki1983,akl1986,Mirlin2006,Janssen1998,Evers2008}. However, recent studies of the artificial Bethe lattice~\cite{Biroli1211,DeLuca2014}, random matrix models~\cite{Kravtsov2015}, and dipolar excitations in 3D random systems~\cite{Deng2016} have revealed finite-width bands of extended non-ergodic states next to the ergodic bands, raising fundamental questions concerning ergodic-to-non-ergodic transitions~\cite{Altshuler2016}. 

Beyond nearest-neighbor hopping breaks the self-duality of the AAH model, and energy-dependent mobility edges appear~\cite{Aubry1980, Ricklund1986, Boers2007,Biddle2010, Biddle2011, Li2017}.  This is the case in shallow lattices, where intermediate regimes with both extended and localized SPS have been predicted~\cite{Boers2007,Biddle2010, Biddle2011,Li2017} and recently observed~\cite{Lueschen2018}, or in zig-zag lattices with next-to-nearest neighbor hopping~\cite{Gadway2018}. Self-duality is also lost when the hopping amplitude decays with the interparticle distance $r$ as $1/r^a$~\cite{Biddle2011,footnote-others}. This is particularly interesting since power-law interactions occur in many systems. Dipole-dipole interactions~($\propto 1/r^3$)  play a crucial role 
for magnetic atoms~\cite{DePaz2013}, polar molecules~\cite{Yan2013}, Rydberg atoms~\cite{Saffman2010}, nitrogen-vacancy centers~\cite{Waldherr2014}, and
nuclear spins in solid-state systems~\cite{Alvarez2015}. Moreover, tunable power-law interactions are achievable for laser-driven ions ($0\le a\le 3$)~\cite{Richerme2014,Jurcevic2014} and for atoms in photonic crystal waveguides~\cite{Hung2016}.  These interactions induce power-law exchange, e.g.~between rotational states in polar molecules~\cite{Yan2013} or hyperfine states in trapped ions~\cite{Richerme2014, Jurcevic2014}, resulting in power-law hopping of excitations. 

In this Letter, we study the SPS of generalized AAH~(GAAH) models with power-law hops. Short-range hops ($a>1$) are characterized 
by a hierarchy of regimes with mobility edges~ (Fig.~\ref{fig:1}). Remarkably, for long-range hops, $a\leq 1$, all SPS are extended, in stark contrast to power-law models with uncorrelated disorder~\cite{Burin1989,Celardo2016,Deng2018}. However, there are finite-width bands of both ergodic and non-ergodic (multifractal) states. We classify these states and show that there is a phase transition at the ergodic-to-multifractal edge, characterized by an abrupt change of fractal dimensions. Moreover, we show that the expansion dynamics of excitations can reveal the presence of mobility and ergodic-to-multifractal edges.



\paragraph{Generalized AAH model.-} We consider pinned particles (with unit filling) at the sites of a deep 1D lattice.
The particles have two internal states $\{\uparrow,\downarrow\}$. Interactions result in power-law exchange between particles. 
A second lattice, incommensurate with the primary one, induces a quasi-disordered variation of the 
energy difference between $\uparrow$ and $\downarrow$~\cite{footnote:difpol}.
The transport of an $\uparrow$ excitation in a sample of $\downarrow$ particles is described by a GAAH model:
\begin{equation}
\!\hat H \!=\! -J\!\sum_{i,j\neq i} \! \frac{1}{|i-j|^a} |i\rangle\langle j| \!+\! \Delta \!\sum_j \!\cos(\beta(2\pi j+\phi)) |j\rangle\langle j|, 
\label{eq:H} 
\end{equation}
where $|j\rangle$ denotes the state in which the excitation is localized at the site $j$, and
$J/|i-j|^a$ is the hopping rate between the sites $i$ and $j$. We set $J=1$ for simplicity.
The quasi-disorder potential is characterized by its strength $\Delta$, the incommesurabilty $\beta$ (the ratio of the period of the primary lattice to the one of the second lattice), and the displacement $\phi$.
For $a\gg 1$, the GAAH model approaches the AAH model~\cite{footnote-SM}. For the latter, all SPS are ergodic for $\Delta<2$, all localized for $\Delta>2$, and all multifractal (extended but non-ergodic) at $\Delta=2$~\cite{Aubry1980,Suslov1982,Wilkinson1984}. 



\paragraph{Determination of the localization properties.-}  
A way of discerning between localized, multifractal, and ergodic SPS, which is especially 
useful for spectra with edges, is given by the analysis of the eigenenergies $E_n$~(indexed in growing energy order), 
and in particular by the even-odd~(odd-even) spacings $s^{e-o}_{n}=E_{2n}-E_{2n-1}$~($s^{o-e}_{n}=E_{2n+1}-E_{2n}$).
Ergodic SPS present a doubly-degenerate spectrum~($s^{o-e}_{n}\simeq 0$)~\cite{Aubry1980}, and hence a gap between $s^{e-o}_{n}$ and $s^{o-e}_{n}$. 
In contrast, for localized SPS both subsets are of the same form, and the gap vanishes. This is illustrated for the AAH model in Fig.~\ref{fig:2}~(a1-a3). 
For the multifractal case~($\Delta=2$) the distribution of both $s^{e-o}_{n}$ and $s^{o-e}_{n}$ is strongly scattered~(Fig.~\ref{fig:2}~(a2)). We also characterize the SPS $|\psi_{n} \rangle=\sum_j \psi_{n}(j) |j\rangle$ by the moments $I_q(n)=\sum_j |\psi_{n}(j)|^{2q}\propto N^{-D_q(q-1)}$, where $D_q$ are the fractal dimensions. Localized states are characterized by $D_q=0$, ergodic extended states by $D_q=1$, while multifractal states have non-trivial $0<D_q<1$~\cite{Wegner1980,Aoki1983,akl1986,Mirlin2006,Janssen1998,Evers2008,DeLuca2014,Deng2016}. As shown below, the study of $D_2$~(obtained from the inverse participation ratio, $I_2$) is particularly useful to characterize transitions at the mobility and ergodic-to-multifractal edges. 
The study of the multifractal spectrum of the SPS and $D_{q>2}$~\cite{DeLuca2014} confirms the classification provided by the level spacing and $D_2$ analyses~\cite{footnote-SM}.



\paragraph{Mobility edge.-} Figure~\ref{fig:1} summarizes our results for $\beta=(\sqrt{5}-1)/2$, but similar physics is found for other values of $\beta$. We assume periodic boundary conditions in our exact-diagonalization calculations, choosing the number of sites $L$ within the Fibonacci series~(up to $L=75025$). 
For $a\gg 1$ we recover the AAH model, and hence 
all SPS are ergodic (AE regime) for $\Delta<2$, or localized (AL regime~\cite{footnote-AL}) for $\Delta>2$.
For finite $a>1$ there is a critical value $\Delta_0(a)$ at which a mobility edge splits ergodic and localized SPS~\cite{footnote1}. 
For $\beta=(\sqrt{5}-1)/2$ we numerically find that states with energies $E_{\beta L\leq n \leq L}$ become localized~(Fig.~\ref{fig:2}~(b1)), whereas those with $E_{n<\beta L}$ remain ergodic. 
This regime, which we call $P_1$, exists up to a critical $\Delta_1(a)$, at which states with energies $E_{\beta^2L<n<\beta L}$ also become localized~(Fig.~\ref{fig:2}~(b2)). The localization transition is observable from the behavior of $D_2$, which in our calculations springs from $1$ to a value that within our numerical accuracy is compatible with $D_2=0$~(Fig.~\ref{fig:2}~(b3)). A sequence of $P_s$ regimes is present for higher $\Delta$ values~(Fig.~\ref{fig:1})~\cite{footnote-LargePs}. In the $P_s$ regime the lowest $\beta^s L$ states are ergodic and the rest are localized.  
The block-like nature of the transitions may be well understood from the analysis of the dispersion of the subbands~\cite{footnote-SM}.  We note that the above mentioned 
particular blocks of states that localize or become multifractal for $\beta=(\sqrt{5}-1)/2$ result from the form of the corresponding bands. Although the overall form of the diagram of Fig.~\ref{fig:1} is maintained for other $\beta$ values, the sizes of the eigenstate blocks, as well as the specific boundaries, $\Delta_s(a)$, of the $P_s$ regimes, depend on the value of $\beta$.



\paragraph{Ergodic-to-multifractal edge.-} Interestingly, the SPS properties radically change for long-range hops~($a\leq 1$). The AE regime extends all  the way down to $a=0$, where it vanishes. 
The sequence of $P_s$ regimes is maintained, but localization is absent, in stark contrast to the case of power-law hopping in the presence of uncorrelated disorder~\cite{Deng2018}. 
In contrast, the spectrum presents an edge between ergodic and multifractal~(extended but non-ergodic) SPS.
Within the $P_s$ regime, the lowest $\beta^s L$ states are ergodic, whereas the rest are multifractal. 
This behavior is illustrated for $a=0.5$ in Figs.~\ref{fig:2}~(c1,c2). When crossing the $P_{s-1}$ to $P_s$ boundary,   $D_2$ jumps from $1$ to $0<D_2<1$ for the states with energies 
$E_{\beta^s L<n<\beta^{s-1}L}$~\cite{footnote-Energies-D2}. This confirms the ergodic-to-multifractal 
character of the transition~(Fig.~\ref{fig:2}~(c3))~\cite{footnote-SM}.



\begin{figure}[t]
\begin{center}
\includegraphics[width =1.0\columnwidth]{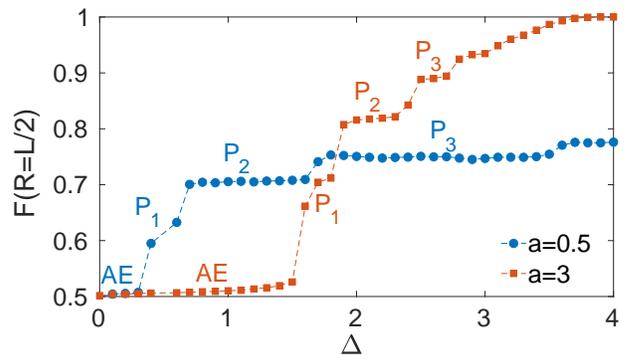}
\caption{(Color online) Survival probability $F(R=L/2)$ in the long-time limit~($Jt=10^4$) for $a=0.5$~(blue circles) and $a=3$~(red squares), for open boundary conditions with $L=987$ sites and $\beta=(\sqrt{5}-1)/2$.  The intermediate $P_s$ regimes lead to a step-wise dependence on $\Delta$.}
\vspace{-0.5cm}
\label{fig:3}
\end{center}
\end{figure}




\begin{figure}[t]
\begin{center}
\includegraphics[width =1.0\columnwidth]{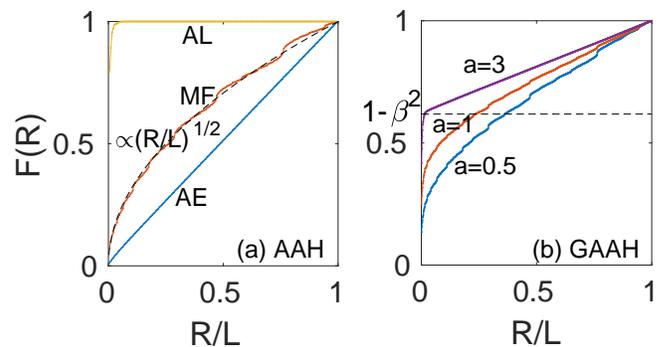}
\caption{(Color online) (a) Long-time survival probability $F(R)$ for the AAH model with $L=100$, for the AE, AL and MF cases, assuming an initially localized excitation at $x=0$  in $L=100$ sites. (b) $F(R)$ for the GAAH model with open boundary conditions and $L=987$ sites for $a=0.5$, $1$ and $3$, within the $P_2$ regime.}
\vspace{-0.5cm}
\label{fig:4}
\end{center}
\end{figure}




\paragraph{Excitation dynamics.-} The nature of the SPS results in a peculiar excitation dynamics. 
We consider all particles $\downarrow$, except an initially localized $\uparrow$ excitation, which for simplicity is placed at the center of a lattice with open boundary conditions. We define the survival probability,  $F(R)$, as the probability of finding the excitation after a given time in a site
within the region $(-R/2,R/2)$. As recently shown for random matrix models~\cite{Kravtsov2012,Torres2018,Santos2018,DeTomasi2018}, $F(R)$ provides a crucial information about the localization properties.
Figure~\ref{fig:3} shows $F(R=L/2)$ as a function of $\Delta$ for $a=3$ and $a=0.5$
for open boundary conditions and $L=987$ sites for long times $t$~($Jt=10^4$), although similar results are found for smaller lattices and 
shorter times. In the AE regime, $F(R)$ vanishes for infinitely large $L$ and long times. For finite $L$ the probability of finding the excitation at a given lattice site is the same for all sites 
and is equal to $1/L$. In contrast,  
the $P_s$ regimes present localized and extended SPS, and hence the 
excitation wavepacket presents a bimodal distribution, partially escaping, and partially remaining localized close to the initial position. As a result $F(R)$ presents a step-like growth when entering the $P_s$ 
regimes (see Fig.3).  

The dynamics in the presence of multifractal SPS 
differs from that of localized and ergodic ones. 
This is best illustrated in the AAH model~(see Fig.~\ref{fig:4}~(a)). 
For sufficiently large $R/L$, for long times, $F(R) \simeq 1$ for $\Delta>2$, $F(R) \simeq R/L$ for $\Delta<2$, and $F(R) \simeq  (R/L)^{1/2}$ for $\Delta=2$. 
The latter reflects the non-ergodic character of the multifractal expansion.
For the GAAH model with $\beta=(\sqrt{5}-1)/2$, in the $P_s$ regime the lowest $\beta^s L$ states remain extended. 
If the rest of the SPS are localized~($a>1$), with a localization length smaller than $R/4L$, then $F(R)$ approaches $F_0(R)=(1-\beta^s)+\beta^s R/L$~($a=3$ in Fig.~\ref{fig:4}(b)). 
However, for $a\leq 1$ there are ergodic and multifractal SPS, and the latter also contribute to the escape probability.
Hence,  for $L\to\infty$ the function $F(R)$ should vanish for all $P_s$ regimes. 
For finite systems, $F(R)$ remains finite, but $F(R) < F_0(R)$ and presents a nonlinear 
dependence~(Fig.~\ref{fig:4}(b))~\cite{footnote-largesystems}. 

\begin{figure}[t]
\begin{center}
\includegraphics[width =1.\columnwidth]{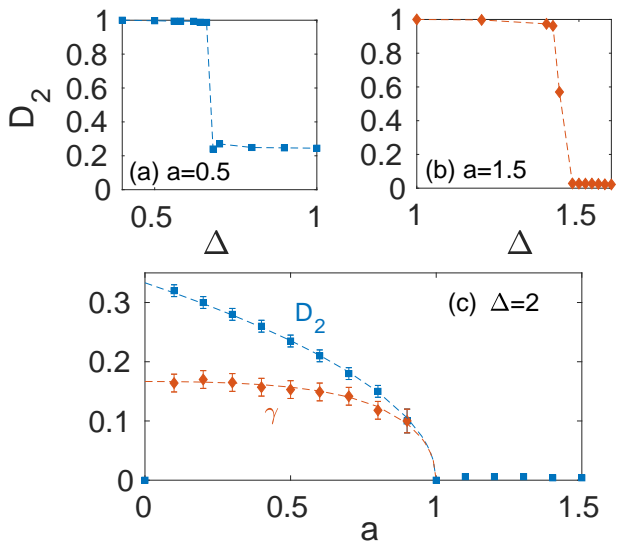}
\vspace*{-0.5cm}
\caption{(Color online) Fractal dimension $D_2$ for $a=0.5$~(a) and $1.5$~(b) as a function of $\Delta$. 
 Figure (c) shows $D_2$ and $\gamma$~(with $l(t)\propto t^{-\gamma}$) as a function of $a$ for $\Delta=2$. 
 Dashed lines depict the fitted relations $D_2(a)\approx \frac{1}{3}(1-a)^{1/2}$ and $\gamma\simeq D_2/(2-a)$. 
See Ref.~\cite{footnote-SM} for a detailed discussion on our calculations of $D_2$ and $\gamma$, as 
well as on the error bars.}
\vspace*{-0.5cm}
\label{fig:5}
\end{center}
\end{figure}

The time dependence of $F(R)$ constitutes as well a clear indicator of the presence of multifractal SPS~\cite{Kravtsov2012,Torres2018,Santos2018,DeTomasi2018}. Figure~\ref{fig:5} 
shows our results for $l(t)=F(R=0,t)=|\langle \psi(t) | \psi(0) \rangle|^2$~(i.e. the Loschmidt echo amplitude, where $\psi(0)$ and $\psi(t)$ are the initial state and its evolved state, respectively). For all cases $l(t)\sim t^{-\gamma}$. 
Fitting our numerical data to this dependence we find that ergodic~(localized) SPS result in $\gamma\simeq 1$~($0$), whereas our numerics reveals that 
the multifractal SPS appearing for $a\leq 1$ result in $\gamma\simeq D_2/(2-a)$~\cite{footnote-SM}. 
The analysis of the excitation dynamics, which can be monitored using spin-resolved quantum microscopes~\cite{Covey2018}, can hence 
reveal not only the structure of intermediate regimes, but also the multifractal nature of the SPS for long-range hops.



\paragraph{Outlook.-} Quasicrystals with power-law hops, $1/r^a$, present non-trivial localization properties. They are characterized by mobility edges for $a>1$, by ergodic-to-multifractal edges for $a\leq 1$, and by the existence of a ladder of intermediate regimes in which SPS blocks become localized or multifractal. 
These properties may be readily tested using expansion experiments. Mobility edges and step-wise dynamics may be experimentally probed for polar molecules pinned in deep bichromatic optical lattices. Powers $0 \leq a\leq 3$ may be directly realized in ions~\cite{Richerme2014,Jurcevic2014}. Hence ion experiments are particularly interesting for the comparative study of mobility versus ergodic-to-multifractal edges. 

Ising-like interactions, which for the case of spin excitations in polar molecules may be induced by an external polarizing electric field, are expected to lead to an intriguing physics including the possibility of a many-body mobility or ergodic-to-non-ergodic edge, due to the interaction-induced coupling between ergodic and localized or multifractal SPS~\cite{Li2015,Modak2015}. 
The presence of non-ergodic SPS bands opens fascinating possibilities for the realization of a bad metal phase~\cite{Pino2016,Pino2017} and for the observation of ergodic to non-ergodic phase transition.



We acknowledge fruitful discussions with V. E. Kravtsov and O. M. Yevtushenko. X. D. and L. S. thank the support of the DFG (SFB 1227 DQ-mat and FOR2247). G. V. S. acknowledges funding from the European Research Council under European Community's Seventh Framework Programme (FP7/2007-2013 Grant Agreement no. 341197). S.S. would like to thank the Institute of Theoretical Physics of the Leibniz University of Hannover for its hospitality.


\begin{thebibliography}{99}

\bibitem{Anderson1958} P.~W.~Anderson, Phys. Rev. {\bf 109}, 1492 (1958).

\bibitem{Abrahams1979}E. Abrahams, P. W. Anderson, D. C. Licciardello, and T. V. Ramakrishnan, Phys. Rev. Lett. {\bf 42}, 673 (1979).

\bibitem{Harper1955} P.~G.~Harper, Proc. Phys. Soc. A {\bf 68}, 874 (1955). 

\bibitem{Azbel1964} M.~Ya.~Azbel, Sov. Phys. JETP {\bf 17}, 665 (1963), Sov. Phys. JETP {\bf 19}, 634 (1964), Phys. Rev. Lett. {\bf 43}, 1954 (1979).

\bibitem{Aubry1980} S.~Aubry and G.~Andr\'e, Ann. Isr. Phys. Soc. {\bf 3}, 133 (1980).

\bibitem{Roati2008} G.~Roati, C.~D'Errico, L.~Fallani, M.~Fattori, C.~Fort, M.~Zaccanti, G.~Modugno, M.~Modugno, and M.~Inguscio, Nature (London) {\bf 453}, 895 (2008).

\bibitem{Deissler2010} B. Deissler, M. Zaccanti, G. Roati, C. D'Errico, M. Fattori, M. Modugno, G. Modugno, and M. Inguscio, Nat. Phys. {\bf 6}, 87 (2010). 

\bibitem{Schreiber2015} M. Schreiber, S. S. Hodgman, P. Bordia, H. P. L\"schen, M. H. Fischer, R. Vosk, E. Altman, U. Schneider, and I. Bloch, Science {\bf 349}, 842 (2015).

\bibitem{Lueschen2017} H. P. L\"uschen, P. Bordia, S. Scherg, F. Alet, E. Altman, U. Schneider, and I. Bloch,  Phys. Rev. Lett. {\bf 119}, 260401 (2017).

\bibitem{Suslov1982} I. M. Suslov, Sov. Phys. JETP {\bf 56}, 612 (1982).

\bibitem{Wilkinson1984} M. Wilkinson, Proc. R. Soc. Lond. A {\bf 391}, 305 (1984).

\bibitem{Wegner1980} F. Wegner, Z. Phys. B 36, 209 (1980).

\bibitem{Aoki1983} H. Aoki, J. Phys. C {\bf 16}, L205 (1983); Phys. Rev. B {\bf 33}, 7310 (1986).

\bibitem{akl1986} B.L. Altshuler, V.E. Kravtsov, and I.V. Lerner, JETP Lett. {\bf 43}, 441 (1986).

\bibitem{Mirlin2006} A.~D.~Mirlin, Y.~V.~Fyodorov, A.~Mildenberger, and F.~Evers, Phys. Rev. Lett. {\bf 97}, 046803 (2006).

\bibitem{Janssen1998} M. Janssen, Phys. Rep. {\bf 295}, 1 (1998).

\bibitem{Evers2008} F.~Evers and A.~D.~Mirlin, Rev. Mod. Phys. {\bf 80}, 1355 (2008).

\bibitem{Biroli1211} G. Biroli, A. Ribeiro-Teixeira, and M. Tarzia, arXiv:1211.7334.

\bibitem{DeLuca2014} A.~De~Luca, B.~L.~Altshuler, V.~E.~Kravtsov, and A.~Scardicchio, Phys. Rev. Lett. {\bf 113}, 046806 (2014).

\bibitem{Kravtsov2015} V.E. Kravtsov, J.M. Khaimovich, E. Cuevas, and M. Amini, New J. Phys. {\bf 17}, 122002 (2015).

\bibitem{Deng2016} X. Deng, B.L. Altshuler, G.V. Shlyapnikov, and L. Santos, Phys. Rev. Lett. {\bf 117}, 020401 (2016).

\bibitem{Altshuler2016} B. L. Altshuler, E. Cuevas, L. B. Ioffe, and V. E. Kravtsov, Phys. Rev. Lett. {\bf 117}, 156601 (2016).

\bibitem{Ricklund1986} R. Riklund, Y. Liu, G. Wahlstr\"om and Z. Zhao-bo, J. Phys. C {\bf 19}, L705 (1986).

\bibitem{Boers2007} D. J. Boers, B. Goedeke, D. Hinrichs, and M. Holthaus, Phys. Rev. A {\bf 75}, 063404 (2007).

\bibitem{Biddle2010} J. Biddle and S. Das Sarma, Phys. Rev. Lett. {\bf 104}, 070601 (2010).

\bibitem{Biddle2011} J. Biddle, D. J. Priour Jr., B. Wang, and S. Das Sarma, Phys. Rev. B {\bf 83}, 075105 (2011).

\bibitem{Li2017} X. Li, X. Li, and S. Das Sarma, Phys. Rev. B {\bf 96}, 085119 (2017).

\bibitem{Lueschen2018} H. P. L\"uschen, S. Scherg, T. Kohlert, M. Schreiber, P. Bordia, X. Li, S. Das Sarma, and I. Bloch, Phys. Rev. Lett. {\bf 120}, 160404 (2018).

\bibitem{Gadway2018} F. A. An, E. J. Meier, and B. Gadway, Phys. Rev. X {\bf 8}, 031045 (2018).

\bibitem{footnote-others} For other functional dependences of the hopping amplitude see Refs.~\cite{Biddle2010,Biddle2011}.
  
\bibitem{DePaz2013} A.~de~Paz, A.~Sharma, A.~Chotia, E.~Mar\'echal, J.~ H.~Huckans, P.~Pedri, L.~Santos, O.~Gorceix, L.~Vernac, and B.~Laburthe-Tolra, Phys. Rev. Lett. {\bf 111}, 185305 (2013).

\bibitem{Yan2013} B.~Yan, S.~A.~Moses, B.~Gadway, J.~P.~Covey, K.~R.~A.~Hazzard, A.~M.~Rey, D.~S.~Jin, and J.~Ye, Nature {\bf 501}, 521 (2013).

\bibitem{Saffman2010} M.~Saffman, T.~G.~Walker, and K.~M\o lmer, Rev. Mod. Phys. {\bf 82}, 2313 (2010).

\bibitem{Waldherr2014} G.~Waldherr, Y.~Wang, S.~Zaiser, M.~Jamali, T.~Schulte-Herbrueggen, H.~Abe, T.~Ohshima, J.~Isoya, P.~Neumann, and J.~Wrachtrup, Nature {\bf 506}, 204 (2014).

\bibitem{Alvarez2015} G.~A.~\'Alvarez, D.~Suter, and R.~Kaiser, Science {\bf 349}, 846 (2016).

\bibitem{Richerme2014} P.~Richerme, Z.-X.~Gong, A.~Lee, C.~Senko, J.~Smith, M.~Foss-Feig, S.~Michalakis, A.~V.~Gorshkov, and C.~Monroe, Nature {\bf 511}, 198 (2014).

\bibitem{Jurcevic2014} P.~Jurcevic, B.~P.~Lanyon, P.~Hauke, C.~Hempel, P.~Zoller, R.~Blatt, and C.~F.~Roos, Nature {\bf 511}, 202 (2014).

\bibitem{Hung2016} C.-L.~Hung, A.~Gonzalez-Tudela, J.~I.~Cirac, and H.~J.~Kimble, PNAS {\bf 113} E4946 (2016).

\bibitem{Burin1989} A. L. Burin and L. A. Maksimov,  JETP Lett. {\bf 50}, 338 (1989).

\bibitem{Celardo2016} G.~L.~Celardo, R.~Kaiser, and F.~Borgonovi, Phys. Rev. B {\bf 94}, 144206 (2016).

\bibitem{Deng2018} X. Deng, V. E. Kravtsov, G. V. Shlyapnikov, and L. Santos, Phys. Rev. Lett. {\bf 120}, 110602 (2018).

\bibitem{footnote:difpol} Differential polarizability results in the quasi-disordered variation of energy between both states. Differential polarizability is irrelevant for the primary lattice, as long as both spin components can be considered as pinned.

\bibitem{footnote-SM} See the Supplementary Material for more details 
on an effective band-dependent tight-binding model that allows for a simple 
understanding of the block-like transitions between $P_s$ regimes, on the analysis of multifractality of single-particle states, on the special case of $a=1$, on the 
relation between inverse participation ratio and survival probability, on the convergence of the GAAH model towards the AAH model when $a\to\infty$, and on details of
localization and ergodic-to-multifractal transitions.

\bibitem{footnote-AL} For a discussion on the existence of the AL regime in the GAAH model see A. M. Garc\'ia-Garc\'ia and E. Cuevas, Phys. Rev. B {\bf 79}, 073104 (2009); C. Monthus, Fractals, doi: /10.1142/S0218348X19500075; arXiv:1706.04099. M. G\"arttner, S. V. Syzranov, A. M. Rey, V. Gurarie, and L. Radzihovsky, Phys. Rev. B {\bf 92}, 041406(R) (2015).

\bibitem{footnote1} A finer analysis, as well as the study of the sub-band dispersion discussed in the Supplementary Material~\cite{footnote-SM}, shows that the transition between $P_s$ regimes actually occurs within a very narrow but finite window of $\Delta$ values, 
in which sub-bands become localized. These sub-band transitions may be as well split into even narrower sub-sub-bands transitions.  
However, for all practical purposes this self-similar set of closely-lying transitions may be considered as a single abrupt jump between $P_s$ regimes. 

\bibitem{footnote-LargePs} For larger $\Delta$, we observe in our numerics a series of $P_s$ regimes, due to the self-similar nature form of the band spectrum. However, for large $s$, the $P_s$ regime is characterized for $\beta=(\sqrt{5}-1)/2$ by a tiny fraction $\beta^s L$ of ergodic states.
Hence, for finite systems eventually the $P_s$ ladder ends into an AL phase. We note however, that in our numerics for $a<2$ we have always found, even for very large $\Delta$ values a finite region of ergodic states when considering larger and larger systems. This explains the form of the diagram in Fig.~\ref{fig:1}. For the $P_{s\gg 1}$ regime the number of ergodic states is tiny and becomes irrelevant for the expansion dynamics.

\bibitem{footnote-Energies-D2} The value of $D_2$ for the multifractal states does not variate significantly either as a function of the 
eigenenergy or of the quasi-disorder strength $\Delta$~\cite{footnote-SM}. 




\bibitem{Kravtsov2012} V. E. Kravtsov, O. M. Yevtushenko, P. Snajberk, and E. Cuevas, Phys. Rev. E {\bf 86}, 021136 (2012).

\bibitem{Torres2018} E. J. Torres-Herrera, Antonio M. Garc\'ia-Garc\'ia, and Lea F. Santos, Phys. Rev. B {\bf 97}, 060303(R) (2018).

\bibitem{Santos2018} L. F. Santos and E. J. Torres-Herrera, arXiv:1803.06012 (2018).

\bibitem{DeTomasi2018} G. De Tomasi, M. Amini, S. Bera, I. M. Khaymovich, and V. E. Kravtsov, arXiv:1805.06472.

\bibitem{footnote-largesystems} The form of $F(R)$ becomes independent of the size $L$ for the case of an ergodic-to-multifractal edge, showing that the states are actually multifractal and not only localized with a large localization length.

\bibitem{Covey2018} J. P. Covey, L. De Marco, O. L. Acevedo, A. M. Rey, and J. Ye, New J. Phys. 20, 043031 (2018).

\bibitem{Li2015} X. Li, S. Ganeshan, J. H. Pixley, and S. Das Sarma, Phys. Rev. Lett. {\bf 115}, 186601 (2015).

\bibitem{Modak2015} R. Modak and S. Mukerjee, Phys. Rev. Lett. {\bf 115}, 230401 (2015).

\bibitem{Pino2016} M. Pino, L. B. Ioffe, and B. L. Altshuler, PNAS, 113, 536 (2016).

\bibitem{Pino2017} M. Pino, V. E. Kravtsov, B. L. Altshuler, and L. B. Ioffe, Phys. Rev. B 96, 214205 (2017). 


\end{thebibliography}
\end{document}


\title{Supplementary material of \\ 
"One-dimensional quasicrystals with power-law hopping"}
\author{X. Deng}
\affiliation{Institut f\"ur Theoretische Physik, Leibniz Universit\"at Hannover, Appelstr. 2, 30167 Hannover, Germany}
\author{S. Ray}
\affiliation{Indian Institute of Science Education and Research, Kolkata, Mohanpur, Nadia 741246, India}
\author{S. Sinha}
\affiliation{Indian Institute of Science Education and Research, Kolkata, Mohanpur, Nadia 741246, India}
\author{G. V. Shlyapnikov}
\affiliation{LPTMS, CNRS, Universite Paris Sud, Universite Paris-Saclay, Orsay 91405, France}
\affiliation{SPEC, CEA, CNRS, Universite Paris-Saclay, CEA Saclay, Gif sur Yvette 91191, France}
\affiliation{Russian Quantum Center, Skolkovo, Moscow 143025, Russia}
\affiliation{\mbox{Van der Waals-Zeeman Institute, University of Amsterdam, Science Park 904, 1098 XH Amsterdam, The Netherlands}}
\affiliation{Wuhan Institute of Physics and Mathematics, Chinese Academy of Sciences, 430071 Wuhan, China}
\author{L. Santos}
\affiliation{Institut f\"ur Theoretische Physik, Leibniz Universit\"at Hannover, Appelstr. 2, 30167 Hannover, Germany}

\date{\today}

\begin{abstract}
We provide further information on an effective band-dependent tight-binding model that allows for a simple 
understanding of the block-like transitions between $P_s$ regimes, on the analysis of multifractality of single-particle states, on the special case of $a=1$, on the 
relation between inverse participation ratio and survival probability, on the convergence of the GAAH model towards the AAH model when $a\to\infty$, and on 
localization and ergodic-to-multifractal transitions.
\end{abstract}

\maketitle




\begin{figure}[t]
\begin{center}
\includegraphics[width =1.0\columnwidth]{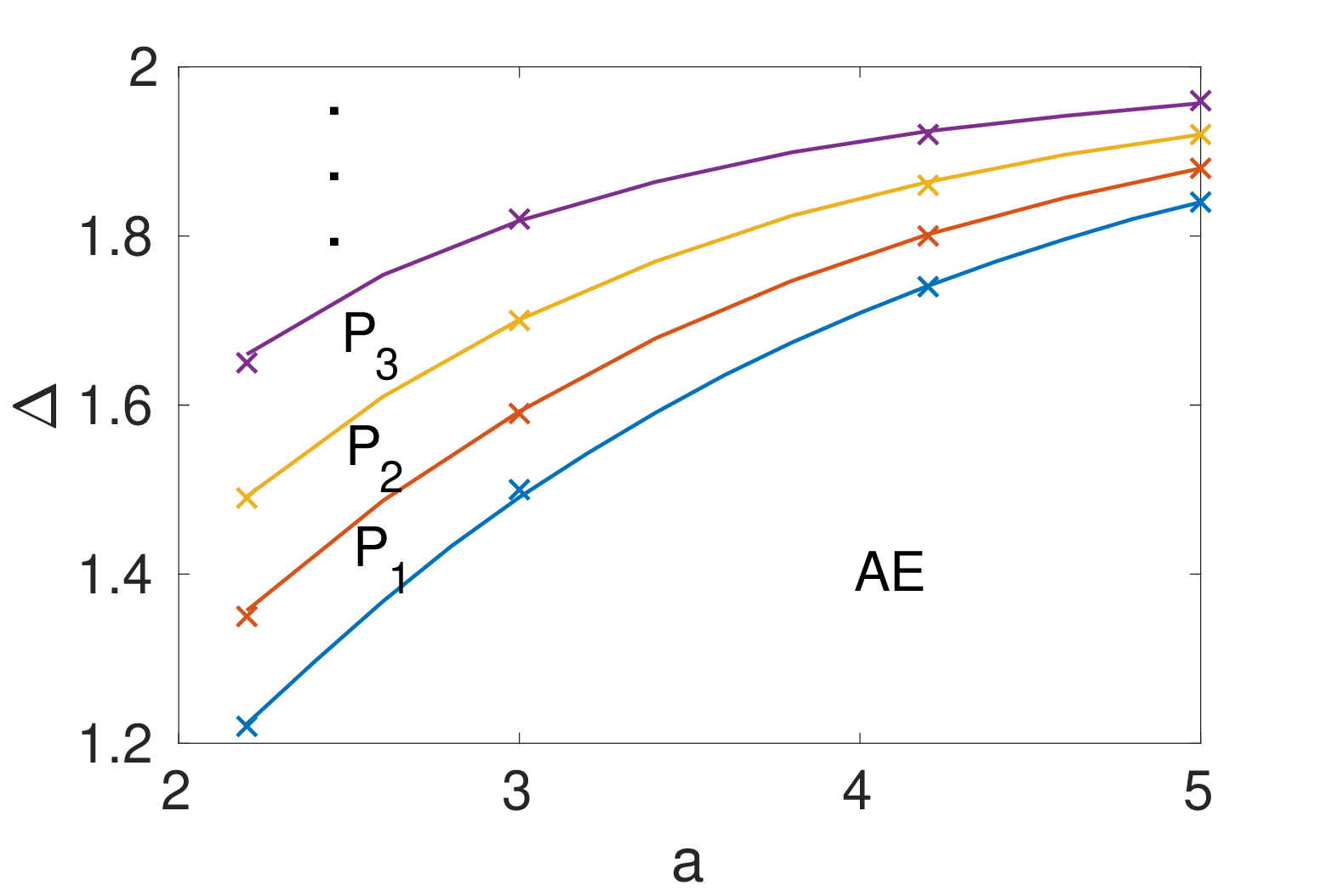}
\vspace*{-0.6cm}
\caption{(Color online) Boundaries of the lowest $P_s$ phases for $\beta=1/(10+\beta)=\sqrt{26}-5$~(crosses) compared to the boundaries calculated from the effective band model, with $\beta=1/10$, discussed in the text~(solid lines).}
\label{fig:SM1}
\vspace{-0.6cm}
\end{center}
\end{figure}


\section{Effective band-dependent tight-binding model}

The abrupt block-like transitions between $P_s$ regimes may be understood from the analysis of the dispersion of the subbands. 
We study the GAAH model:
\begin{eqnarray}
H&=&-J\sum_j \sum_{s=1}^\infty \frac{1}{s^a} \left (c_j^\dag c_{j+s} +{\mathrm H. c. } \right ) \nonumber \\
&-& \Delta \sum_j \cos \left ( \beta (2\pi j + \phi) \right )c_j^\dag c_j,
\label{eq:H}
\end{eqnarray}
We consider $\beta = \frac{1}{p+\beta_1}$, with integer $p\gg 1$, and $\beta_1\ll 1$. We consider first the case $\beta=1/p$. In that case the system presents a period $p$, 
and hence the tight-binding band splits into $p$ subbands. By solving the eigensystem 
\begin{eqnarray}
E \psi(j) &=& -J\sum_j \sum_{r=1}^\infty \frac{1}{r^a} \left ( e^{i\kappa r/n} \psi(j-r) + {\mathrm c.c.} \right ) \nonumber \\
&-&\Delta \sum_j \cos \left ( \beta (2\pi j + \phi) \right ) \psi(j),
\end{eqnarray}
we obtain the dispersion of the subbands as a function of the quasimomentum $-\pi < \kappa \leq \pi$, and the phase shift $-\pi<\phi\leq \pi$. Like in renormalization group calculations for the AAH model~\cite{Suslov1982,Wilkinson1984,Szabo2018},  our numerical calculations show that the $\kappa$ and $\phi$ dependences remain uncoupled, and the subband dispersion acquires the form:
\begin{equation}
E_m(\kappa,\phi) = E_{m,0} - 2 J_{m} {\cal R} \left ( Li_{a_{m}}(e^{i\kappa}) \right ) - \Delta_m \cos(\phi)
\end{equation}
where $m$ denotes the band index, $E_{m,0}$ is the central energy of the $m$-th band, $Li_a(x)=\sum_{s=1}^\infty \frac{x^s}{s^a}$ is the polylogarithm function,
 and ${\cal R}$ indicates the real part. The quasimomentum dependence is the one that is expected for a model with power-law hopping with power $a_m$.  Indeed, introducing the transformation $c_j=\sum_\kappa e^{i\kappa j}b_\kappa$, the hopping term in Eq.~\eqref{eq:H} becomes 
 $-2J\sum_{\kappa} {\cal R}(Li_a(e^{i\kappa})) b_\kappa^\dag b_\kappa$. For $a\gg 1$, ${\cal R}(Li_a(e^{i\kappa}))\simeq \cos(\kappa)$ 
 and the dispersion of a tight-binding model with nearest-neighbor hopping is retrieved.
Hence, the subbands present the dispersion expected for a power-law Hamiltonian, but with regularized band-dependent hopping amplitude $J_m$ and power $a_m$.  In this effective model, the Wannier functions are those of a superlattice with a unit cell which is $p$ times larger than the original one. 
Since $\beta=1/(p+\beta_1)$, there is a slowly varying $\phi=2\pi\beta_1 x$ in the new lattice which leads to a quasi-disorder term with an effective band-dependent strength $\Delta_m$. The effective Hamiltonian for the $m$-th band is hence:
\begin{eqnarray}
H_m&=&-J_m\sum_j \sum_{r=1}^\infty \frac{1}{r^{a_m}} \left (c_j^\dag c_{j+r} + {\mathrm H. c. } \right )\nonumber \\ 
&-&\Delta_m \sum_j \cos \left ( \beta_1 (2\pi j + \phi') \right ) c_j^\dag c_j,
\label{eq:Hm}
\end{eqnarray}
with an effective displacement $\phi'$. We consider below $p=10$, but similar reasonings apply as long as $p$ is sufficiently large.
The evaluation of the most energetic subbands for $a>2$ shows that they present a dispersion with $a_m>6$. For these $a_m$ values, the $\kappa$ dependence of the dispersion is practically indistinguishable from a cosine, $-2J_m\cos(\kappa)$, i.e. the dependence of a model with nearest-neighbor hopping. As a result, the subband behaves effectively as an AAH model~\cite{footnote-2}. Hence, when $\Delta_m/J_m=2$ we expect that the whole subband localizes. Since the relation between $\Delta$ and $\Delta_m/J_m$ is band-dependent, the different subbands localize for different values 
of $\Delta$, starting with the most energetic band $m=p$. This results in the hierarchy of $P_s$ phases discussed in the main text. The $P_s$ phase is characterized by the localization of the highest $s$ bands, and hence the mobility edge would be at a fraction $1-s/p$ of the whole spectrum. For $s/p \ll 1$, the edge may be approximated as $(1-\beta)^s$, which is the numerically observed mobility edge for $\beta=1/(p+\beta)$, with $p>1$. 

From the effective model discussed above we can evaluate the boundaries of the $P_s$ phases. We depict in Fig~\ref{fig:SM1} the case of $\beta=1/(10+\beta)=\sqrt{26}-5$, showing that there is an excellent agreement between the results obtained from the spectral analysis of the GAAH model, and those obtained from the effective band model. The reasoning is however significantly more involved for $a<2$ or for the lowest subbands~(i.e. when approaching the 
boundary of the AL phase), since the subbands retain a significant power law character, i.e. the $\kappa$ dependence of the subband dispersion cannot be well approximated by a cosine. In that case a more careful renormalization analysis, beyond the scope of this paper, is necessary to describe the $P_s$ phases.





\begin{figure}[t]
\begin{center}
\includegraphics[width =1.0\columnwidth]{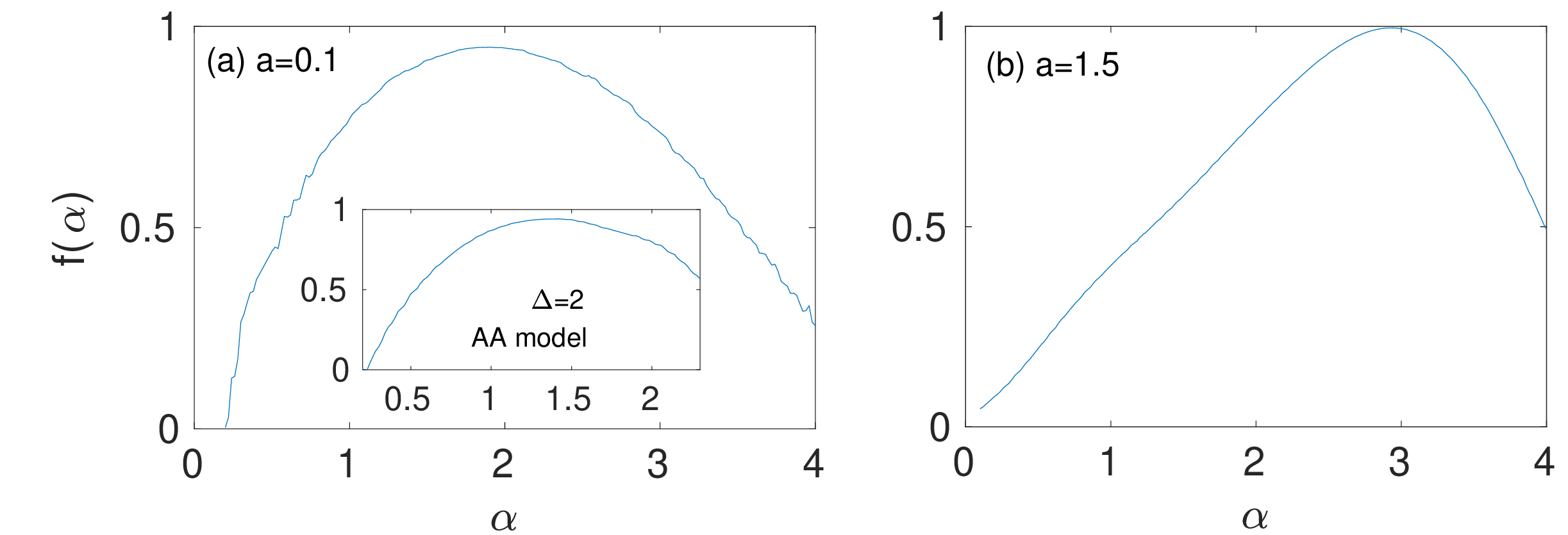}
\caption{(Color online) Multifractal spectrum for states with energies $E_{\beta^2 L<n\le  L}$  
for $a=0.1$~(a) and $1.5$~(b) in the $P_2$ regime. The inset in Fig.(a) shows the results for the AAH model 
with $\Delta=2$. For $a=1.5$ $f(\alpha)=k\alpha$ at small $\alpha$, 
as expected for localized states, whereas both for $a=0.1$ and the critical AAH model $f(\alpha)$ presents the form expected for multifractal states~\cite{DeLuca2014,Deng2016}. 
The MFS is obtained by extrapolating our results (for up to $L=75025$ sites) to an infinite system, employing the technique of Refs.~\cite{DeLuca2014,Deng2016}. 
Convergence is however slow, especially for $a=0.1$, explaining the slightly incorrect determination of the MFS maximum, which 
should be $1$.}
\label{fig:SM2}
\vspace{-0.5cm}
\end{center}
\end{figure}


\section{Multifractality}

In the main text we have monitored the localized, ergodic, or multifractal character of the single-particle states~(SPS) by means of the 
analysis of $D_2$ and the level spacings. As briefly discussed in this section, we have complemented this study 
with the analysis of the multifractal spectrum~(MFS) and fractal dimensions $D_{q>2}$.

For a given eigenstate $|\psi_n \rangle=\sum_j \psi_n(j) |j\rangle$, we may evaluate the moments $I_q(n)=\sum_j |\psi_n(j)|^{2q}\propto L^{-\tau(q)}$, with $L$ the number of sites.
The Legendre transform $\tau(q)=q\alpha-f(\alpha)$ defines the MFS $f(\alpha)$, which
characterizes the Hausdorff dimension of the manifold of sites where $|\psi_{n}(j)|^2=L^{-\alpha}$~\cite{Evers2008,DeLuca2014,Deng2016}. 
Normalization, $\sum_{j}|\psi_{n}(j)|^2=1$, requires $\alpha\geq 0$.  The MFS is evaluated as discussed in Refs.~\cite{DeLuca2014,Deng2016}.

Figure~\ref{fig:SM2} shows our results for $\beta=(\sqrt{5}-1)/2$ for $a=0.1$ and $1.5$ evaluated for values of $\Delta$ within the $P_2$ phase. We monitor 
the eigenstates with energies $E_{\beta^2 L<n\le L}$. In the $P_2$ phase those states are localized for $a>1$ and multifractal for $a<1$. 
Algebraically localized states, $|\psi_{n}(j)|^2\propto 1/|j-j_{0}|^{\gamma}$, present a triangular MFS $f(\alpha)$, 
where $f(\alpha)=k\alpha$ for $0<\alpha<1/k$ with $k=1/\gamma$~\cite{DeLuca2014,Deng2016}. We obtain such a dependence for $a=1.5$. 
On the contrary, for $a=0.1$ the MFS presents the characteristic dependence of multifractal states~\cite{DeLuca2014,Deng2016}, in particular the expected symmetry $f(1+\alpha) = f(1-\alpha)+\alpha$~\cite{Mirlin2006}.

\begin{figure}[t]
\begin{center}
\includegraphics[width =0.7\columnwidth]{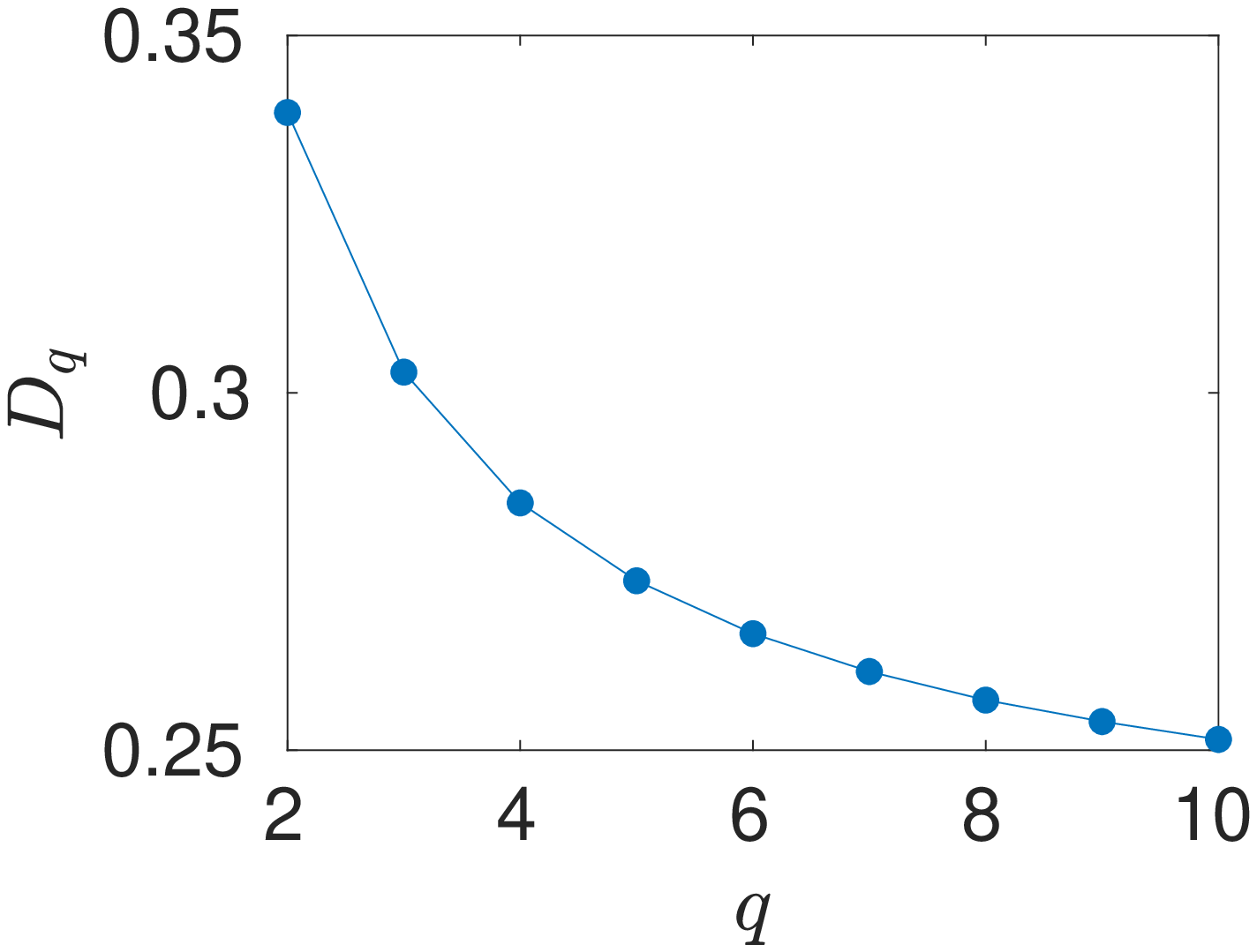}
\caption{(Color online) Fractal dimensions $D_q$ for $\beta=(\sqrt{5}-1)/2$,  $a=0.1$ and $\Delta=2$ for the state with energy $E_{\beta^4 L+1}=-0.9096$, which is multifractal.}
\vspace{-0.5cm}
\label{fig:SM3}
\end{center}
\end{figure}

The multifractal character of the non-ergodic states for $a<1$ is further proved by the analysis of fractal dimensions $D_{q>2}$. Ergodic~(localized) states are characterized by $D_q=1$~($0$) for all $q$, whereas the
non-ergodic states for $a<1$ are characterized by a $q$-dependent $D_q$, as expected for multifractal states~(see Fig.~\ref{fig:SM3}).



\section{Critical line at $a=1$}
We discuss at this point in some more detail the case of $a=1$, since its numerical evaluation is particularly involved.
We may understand this case following the ideas of Ref. \cite{Monthus2018}. In the limit of $J=0$ all eigenstates $|\psi^{(0)}_n\rangle=|n\rangle$ are localized, and the corresponding eigenvalue is given by the on-site quasi-periodic potential, $E_n$. For finite hopping $J/\Delta\ll 1$, the eigenstates may be evaluated to first-order in perturbation theory, 
 $|\psi^{(1)}_n\rangle\simeq |n\rangle+\sum_{n' \ne n}|n'\rangle \langle{n'}|\psi^{(1)}_n\rangle$, with $|\langle{n'}|\psi^{(1)}_n\rangle|^2=|\frac{\frac{1}{|n-n'|^a}}{E_n-E_{n'}}|^2$. The hopping radius scales as $1/L^a$, whereas the level spacings do it as $1/L$. Hence, when $a>1$ the wavefunction is localized, while $a=1$ is critical.
As a result, for $a=1$ the logarithmic correction, $I_q = L^{-\tau(q)}+b(\ln{L})^{\eta(q)}$, plays an important role~\cite{Monthus2018}, and must be carefully considered in our fitting of the inverse participation ratio.
For $a\to 1$ and a large number of sample states, we obtain the best fit of our numerical results for $\eta(2)\approx -1$, which is consistent with Ref. \cite{Monthus2018}, and $\tau(2)=0$~(i.e. $D_2=0$).  
In contrast, for $a=0.8$ we find $b\simeq 0$, and a finite $0<D_2<1$, characteristic of multifractal states.



\section{Inverse participation ratio and survival probability}

\begin{figure}[t]
\begin{center}
\includegraphics[width =0.7\columnwidth]{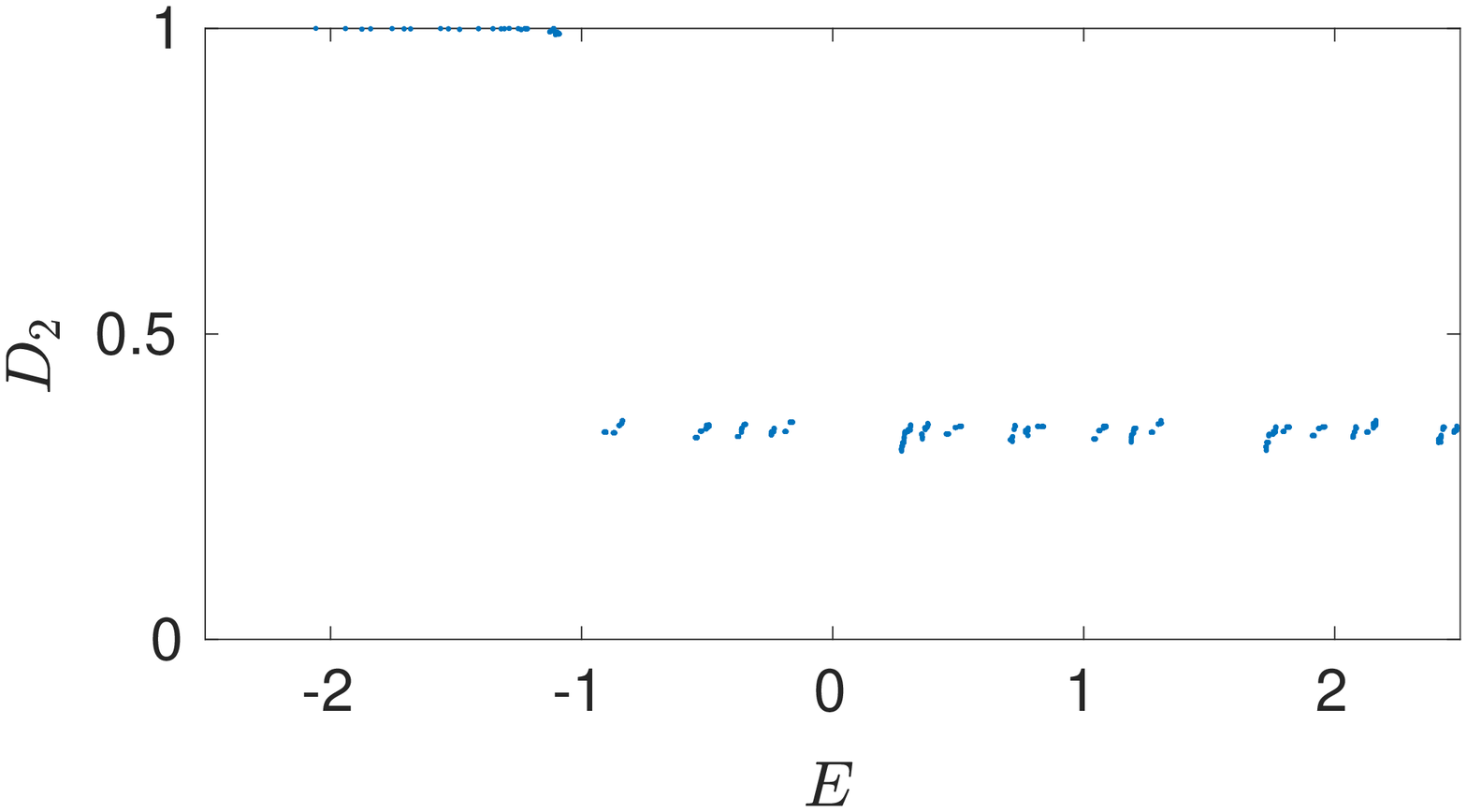}
\caption{(Color online) $D_{2}$ as a function of the energy (with $J=1$) for $\beta=(\sqrt{5}-1)/2$, $a=0.1$, and $\Delta=2$. 
$D_{2}$ do not variate significantly within the family of multifractal states, characterized by $0<D_2<1$.}
\label{fig:SM4}
\end{center}
\end{figure}

The value of $D_2$ may fluctuate within the block of multifractal states. As illustrated in Fig.~\ref{fig:SM4}, we have observed that for $a<1$ the fractal dimension $D_2$ 
does not present a dependence on the energy~(the error bars in Fig. 5(c) of the main text indicate this uncertainty). 
Only in the vicinity of $a=1$ fluctuations of $D_2$ for different multifractal eigenstates become significant. 
This is, on the other hand, expected due to the above mentioned critical nature of the $a=1$ case.
For the determination of $D_2$ close to $a=1$ we evaluate it with the highest accuracy over the multifractal states.
From our numerics, we obtain the best fit $D_2(a)\approx \frac{1}{3}(1-a)^{1/2}$ for $0<a\leq 1$~(for $a=1$, as mentioned above, we obtain 
$I_2\propto 1/\ln L$). Furthermore, in our numerics, for $a<1$ we do not find any localized state up to very large quasi-disorder strengths $\Delta=10$. Indeed, we have checked 
that $D_2$ has no significant $\Delta$ dependence when evaluated for a given $a<1$.



The survival probability for an infinite system is expected to decay with a power law $l(t)\propto t^{-\gamma}$ for $t\to\infty$~\cite{Ketzmerick1992}. 
We have evaluated the exponent $\gamma$ by averaging the function $l(t)$ for up to $100$ different realizations of 
the displacement $\phi$~(we have checked that averaging over further realizations does not change the results) for lattices with up to $L=28657$ sites with open boundary conditions. 
After an initial stage characterized by an exponential decay of $l(t)$, the Loschmidt amplitude is well fitted as $l(t)\sim t^{-\gamma}$.
The observed small oscillations, that we associate to spectral rigidity~\cite{DeTomasi2018}, do not compromise the fitting. We have checked this
by considering the survival probability in a small but finite central region, $F(R=10,t)$. This averaging removes the small oscillations, leaving the bare overall decay $t^{-\gamma}$. 
Fitting $F(R=10,t)$ provides almost identical results as those obtained with the bare $l(t)$. Hence, in the numerical results depicted in the main text we have employed the fit of $l(t)$. 
The small error bars in the determination of $\gamma$, depicted in Fig. 5(c) of the main text, result from the uncertainty of this fitting procedure.

For the AAH model, we obtain $\gamma=1$, $0$, and $\simeq 0.26$~(after averaging over disorder samples) for $\Delta<2$~(ergodic), $\Delta>2$~(localized), and $\Delta=2$~(multifractal), respectively~\cite{Zhong1995}. 
For the GAAH model the long-time evolution of the survival probability is again well fitted by a power-law dependence. In the AE~(AL) regime, we recover $\gamma=1$~($0$).
In the $P_s$ regimes, for $a>1$ the long-time evolution is dominated by the localized states, and hence $\gamma=0$. 
On the contrary, for $a<1$ the long-time dependence of the survival probability is dominated by the (slowly expanding) multifractal states.
We hence obtain $0<\gamma<1$, with a best fit $\gamma\simeq D_2/(2-a)$~(Fig. 5(c) of the main text).
As for the dependence of $D_2$ on the system size, the long-time dynamics of $l(t)$ is also peculiar for $a=1$, for which we observe 
a decay with a power law in $\ln t$.



\section{Convergence of the GAAH model towards the AAH model for $a\to\infty$}
As discussed in the main text, for $a>1$ there are two distinct blocks of eigenstates within the $P_s$ regimes: ergodic and localized states, which for 
$\beta=(\sqrt{5}-1)/2$ are, respectively, those with energies $E_{n\leq\beta^s L}$ and $E_{n>\beta^s L}$. On the other hand, it is known that for the AAH model all 
eigenstates remain ergodic~(localized) for $\Delta<2$~($\Delta>2$), whereas for $\Delta=2$ all states are multifractal. 
As shown in Fig. 1 of the main text, for $a\gg 1$ the $P_s$ regions squeeze into the vicinity of $\Delta=2$. 
However, the question remains on how the GAAH model converges towards the AAH model when $a \to \infty$, i.e. how the multifractal nature of $\Delta=2$ in the AAH model 
is retrieved.

\begin{figure}[t]
\begin{center}
\includegraphics[width =0.7\columnwidth]{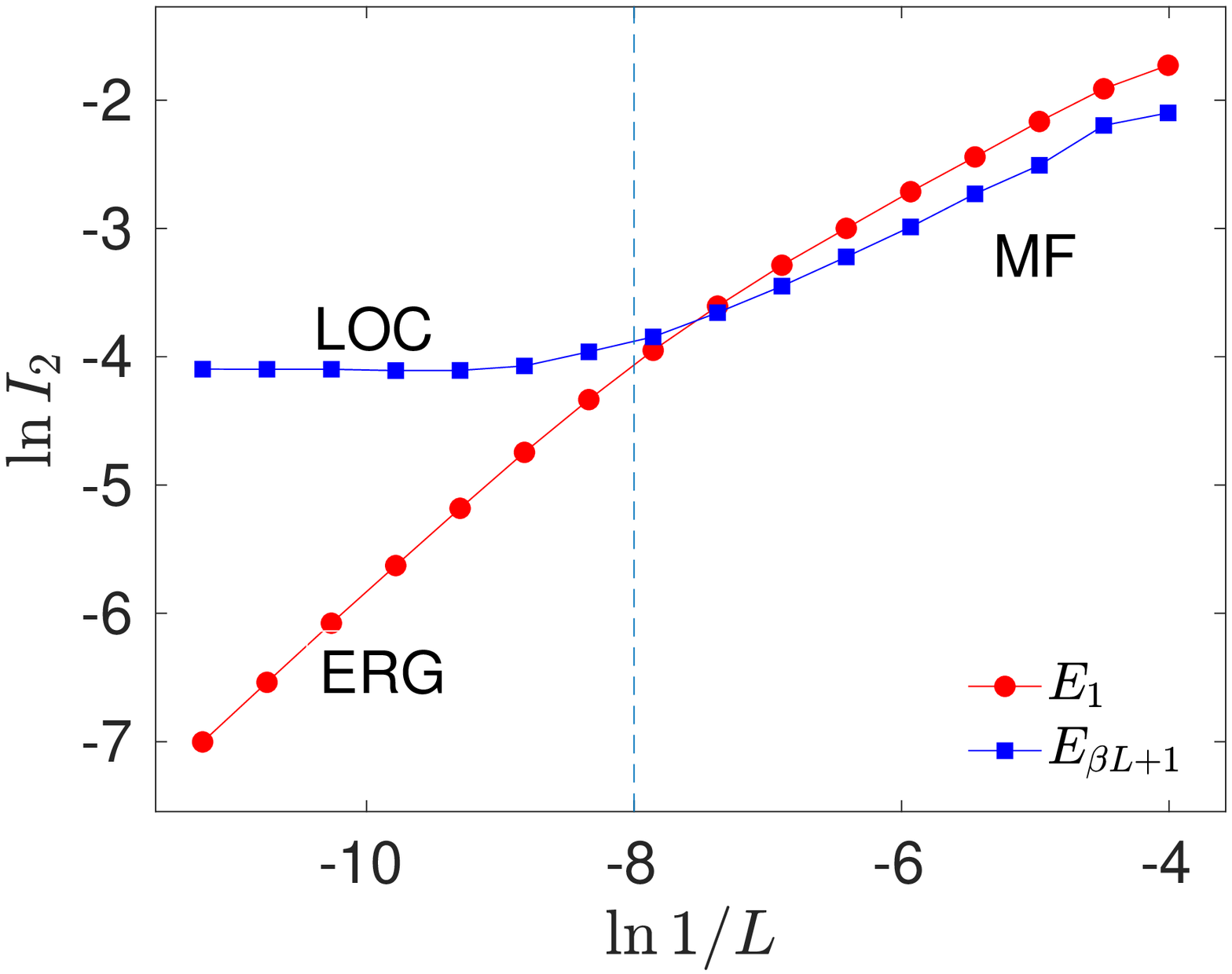}
\caption{(Color online) Inverse participation ratio $I_2$ for $a=12$ as a function of the inverse system size $1/L$ for states with energies $E_{1}$ and $E_{\beta L}$. 
For sufficiently large systems, these states reveal their ergodic~(ERG) and localized~(LOC) nature, respectively. 
In contrast, for small-enough $L$, such that $a\gg \log_2 L$,  the states acquire a multifractal~(MF) character, as expected for the AAH model. 
The dashed vertical line indicates $L=2^a$.}
\label{fig:SM5}
\end{center}
\end{figure}

Figure~\ref{fig:SM5} illustrates this convergence. For $\Delta=2$ within the $P_2$ regime, we evaluate the states 
with eigenenergy $E_n$, at indices $n=1$~(ergodic) and $\beta L$~(localized), for various values of $a$. 
Multifractal states appear in the thermodynamic limit only for $a\to\infty$. 
However, for finite-size systems there are multifractal states for a sufficiently large $a$. This may be understood from the comparison of the energy
 difference between two neighboring levels, $s_L\propto 1/L$, with the next-to-nearest-neighbor hopping, $J_2 \propto 1/2^a$. 
 If $J_2 \ll s_L$, the GAAH model may be approximated by the AAH model and multifractal states appear for $\Delta=2$. 
However, when $J_2 \gg s_L$, the GAAH model has not yet converged to the AAH one, and the system presents only ergodic and localized states. 
Hence, only when  $a \gg log_2L$ the well-known multifractality for $\Delta=2$ in the AAH model is retrieved.
These unusual delocalized states which are ergodic at large scales but non-ergodic at small scales are reminiscent of the scaling of Anderson transition in random graphs~\cite{GarciaMata2017}. 



\begin{figure}[t]
\begin{center}
  \includegraphics[width =0.7\columnwidth]{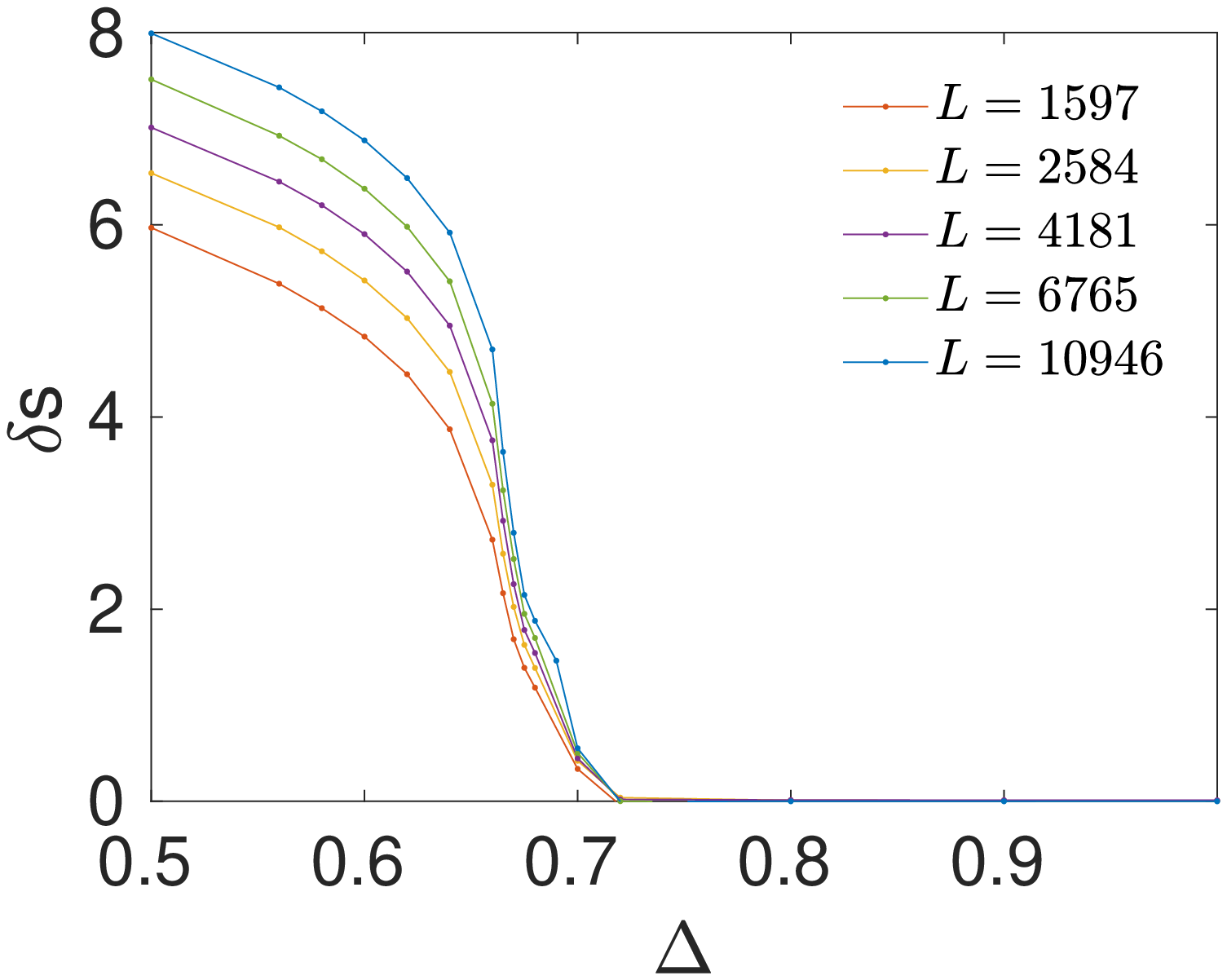}
\caption{(Color online) Level gap $\delta s$ for $E_{\beta^2 L < n \le \beta L}$ and $a=0.5$ as a function of $\Delta$ for different system sizes $L$.}
\label{fig:SM6}
\end{center}
\end{figure}


\section{Localization and ergodic-to-multifractal transitions}

In this section we briefly discuss the transition of eigenstates from ergodic to localized~(multifractal) when increasing $\Delta$ for $a>1$~($<1$). 
We have analyzed these transitions for systems sizes up to $L=75025$ and with periodic boundary conditions, performing a finite-size scaling 
of the transition point. This analysis is illustrated in Fig.~\ref{fig:SM6}, where we consider the case of $a=0.5$, and $0.5<\Delta<1$. In this regime 
the fractal dimension $D_2$ jumps from $1$ to  $\approx 0.24$ for the eigenstates $E_{\beta^2<n\le \beta L}$~(boundary between the $P_1$ and $P_2$ regimes). 
Using the odd-even and even-odd energy spacings introduced in the main text, we define the level gap as $\delta s = \ln(s^{o-e})-\ln(s^{e-o})$. 
We observe that $\delta{s}\propto \ln{L}$. Therefore, using this relation we can extrapolate it to the thermodynamic limit. 
In general, $\delta s$ is finite for ergodic states and vanishes for non-ergodic ones. Figure~\ref{fig:SM6} hence reveals the transition between ergodic and multifractal states.
We obtain similar results when analyzing the localization transition for $a>1$. These results show that the boundaries in Fig. 1 of the main text have very little dependence on the system size.
